# Impact force and spreading characteristics of droplet impact on cylindrical surfaces

Mengqi Ye (叶孟琪)[a], Tianyou Wang (王天友)[a,b], Zhizhao Che (车志钊)[a,b,*]

[a] State Key Laboratory of Engines, Tianjin University, Tianjin, 300350, China

[b] National Industry-Education Platform of Energy Storage, Tianjin University, Tianjin, 300350, China

*Corresponding author: chezhizhao@tju.edu.cn

**Abstract**

Droplet impact phenomena are ubiquitous in both nature and industry. Existing studies of droplet impact have focused on the kinematics of droplet impact on flat surfaces, whereas research on cylindrical surfaces remains relatively limited, particularly from a force-based perspective. Here, droplet impact on cylindrical surfaces is studied by numerical simulation, with particular attention to the spreading behavior and impact force acting on the wall. In the deposition mode, a single peak appears in the impact force curve, which corresponds to the rapid transfer of the droplet's initial momentum. In the rebound mode, two distinct peaks are observed, and the second peak arises from the reaction force during the retraction process. Increasing the surface wettability causes the asymmetry coefficient, the ratio of the maximum spreading lengths in the azimuthal and axial directions of the cylinder, to first decrease and then gradually approach a constant, while the effect of the surface wettability on the initial impact force is negligible. As the Weber number $We$ increases, the first dimensionless peak of the impact force $F_{p1}^*$ approaches a constant, and the relationship can be expressed as $F_{p1}^* = \beta_1 + \beta_2 We^{-1}$ (where $\beta_1$ and $\beta_2$ are constants). The dimensionless maximum spreading area, dimensionless maximum spreading length, dimensionless maximum spreading angle, and asymmetry coefficient all exhibit power-law relationships with the Weber and Ohnesorge numbers. Furthermore, an increase in the diameter ratio of the cylinder and the droplet leads to a reduction in the asymmetry coefficient and an increase in the first dimensionless peak of the impact force.

## I. Introduction

The impact of a droplet on a solid surface is a complex multiphase flow phenomenon that occurs widely in nature and industrial applications[1], such as pesticide spraying[2], spray cooling[3], fog collection[4], and plasma spraying[5]. This process involves many fascinating physical phenomena and complex two-phase flow mechanisms, which have attracted extensive attention from researchers[6-8]. The properties of the droplet (viscosity, density, and surface tension), the surface (wettability, microstructure, and roughness), and the surrounding environment (temperature, electric field, and pressure) all affect the impact behavior during droplet impact[6]. The advancement of computational fluid dynamics and high-speed photography has allowed researchers to study the influencing aspects of droplet impact processes in great detail.

In studies of droplet impact on flat surfaces, the spreading behavior during impact is crucial in many applications and therefore has grown into the primary focus of research. For instance, the spreading length determines the area over which the droplet spreads, and this area directly affects the heat transfer between the droplet and the wall[9]. The relationship between the impact parameters and the maximum spreading length can be obtained by balancing the inertial, viscous, and capillary forces[10-13]. The spreading factor $\xi$, defined as the ratio of the maximum spreading radius to the droplet radius, is typically used to characterize the spreading capability of a droplet. The spreading factor under the viscous regime is a function of the Reynolds number $Re$ and is dependent on both the inertial and viscous forces $\xi \sim Re^{1/5}$ [14] ($Re = \rho_l v_i D_d / \mu_l$, where $\rho_l$, $v_i$, $D_d$, and $\mu_l$ are used to represent the droplet density, impact velocity, droplet diameter, and dynamic viscosity, respectively). While under the inertial regime, the spreading factor may be written as a function of the Weber number $We$ and is controlled by the capillary and inertial forces $\xi \propto We^{1/4}$ [10] ($We = \rho_l v_i^2 D_d / \sigma$, where $\sigma$ is the surface tension coefficient). The wettability of the surface also has major effects on the spreading behavior and droplet impact dynamics. Surfaces with varying wettability can result in various effects, including deposition, rebound, or splashing[15]. When the Weber number is relatively low (< 200), the maximum spreading length is strongly affected by surface wettability, and hydrophilic surfaces promote spreading. While at high Weber numbers (> 200), the influence of surface wettability on the maximum spreading length becomes negligible[16].

Curved surfaces are more common than flat surfaces in practical applications and daily life. In recent years, researchers have investigated the impact of droplets on curved surfaces, primarily focusing on spheres and cylinders[17-20]. When a droplet impacts a slightly curved spherical surface, it typically undergoes successive stages of spreading, stagnation, retraction, and oscillation. A higher Weber number or a smaller contact angle leads to an increased spreading length. Moreover, the spreading length is also affected by the diameter ratio, which is defined as the ratio between the sphere diameter $D_p$ and the droplet diameter $D_d$. When the diameter ratio is greater than the critical value, the spreading length becomes nearly constant[21]. For a constant diameter ratio, droplet spreading is often weaker on concave surfaces than on convex ones[22]. When droplets impact soft spherical surfaces, the effects of elastic modulus and diameter ratio on spreading behavior are relatively limited. Energy dissipation caused by substrate deformation enhances the retraction behavior of droplets impacting highly curved soft surfaces[23].

Compared with flat and spherical surfaces, the impact of a droplet on a cylinder exhibits asymmetric spreading, showing different behaviors in the axial and azimuthal directions, and the impact dynamics become more complex. The contact time for a droplet impacting a hydrophobic flat wall remains nearly unchanged regardless of the impact velocity[24]. However, when impacting a hydrophobic cylinder with a large diameter, the geometric anisotropy of the cylinder significantly shortens the contact time compared with that on a flat wall, and both the impact velocity and the diameter ratio affect the contact time[25]. The dynamic characteristics of droplets impacting cylindrical surfaces show considerable distinctions compared with impacts on flat or spherical surfaces. The impact dynamics can be classified into two typical models: an asymmetric rebound and a stretching breakup, both primarily controlled by $We$ and the diameter ratio[26]. An increase in impact velocity facilitates greater droplet spreading, while the difference between the maximum spreading extents along the azimuthal and axial directions becomes more significant at higher velocities[27]. When droplets impact a thin cylinder (which is usually called a fiber, where the diameter of the cylinder is smaller than the capillary length $\sqrt{\sigma/(\rho_l g)}$, $g$ is the gravitational acceleration), three impact outcomes are observed: capturing, single drop falling, and splitting[28, 29]. In this case, the capture efficiency of the fiber becomes an important parameter of interest for researchers[30, 31]. In addition, studies on eccentric droplet impacts on hydrophobic cylinders have shown two main modes, asymmetric stretched and stretched rebound, and both the eccentricity and the impact velocity contribute to the reduction of the contact time[18].

In studies of droplet impact, researchers mainly focused on the kinematic characteristics of the droplet, such as morphology, maximum spreading length, and contact time, whereas the dynamic characteristics such as impact force, drag force, and stress distribution, have only recently begun to attract significant attention[32-41]. The force generated by the impact of a droplet on a solid wall may induce surface degradation, leading to undesirable effects in both the environment and industrial production, such as soil erosion[42, 43] and the wear of turbine blades[44] and wind turbine blades[45]. However, current understanding of droplet impact erosion in natural and industrial processes remains at an early stage. Early experimental studies quantified the impact force of droplets impacting hydrophilic surfaces by means of piezoelectric sensors. The findings showed that the impact force rapidly reaches a peak shortly after impact and then gradually decays[33, 46, 47]. The impact force behavior can be categorized into three zones depending on the Reynolds number: viscous-dominated, inertia-dominated, and transition zones[32]. The temporal evolution of the impact force exhibits self-similar behavior, with an exponential decay during the attenuation stage in the inertia-dominated zone. The force profiles are independent of *We* but dependent on *Re*, and at high *Re*, it is inertia dominated and depends on the liquid density, impact velocity, and droplet diameter[35]. Compared with hydrophilic surfaces, droplet impact on hydrophobic surfaces exhibits two distinct peaks in the impact force profile. The first peak is similar to that observed for hydrophilic surfaces, while the second peak occurs just before droplet rebound. The regimes of impact force evolution can also be classified according to the Weber number into four regions: capillary, singular jet, inertial, and splashing region[36, 37, 39].

In summary, the majority of earlier studies have emphasized the kinematic aspects of droplet behavior during impact on a flat wall, while research on droplet impact on cylinder remains relatively limited, especially on the dynamic characteristics. Therefore, this study employs numerical simulations to analyze both the kinematic and dynamic behaviors of droplets impacting cylinders. The analysis centers on the spreading dimensions, impact force, and characteristic times associated with these phenomena. Furthermore, the effects of key parameters, including surface wettability, *We*, *Oh*, and diameter ratio, on the impact process are studied. Quantitative relationships are established between the dimensionless maximum spreading area, maximum spreading length, maximum spreading angle, impact force, and asymmetry coefficient with respect to *We* and *Oh*.

## II. Numerical method

The impact of a droplet on a solid surface is a typical multiphase flow process. In this study, the gas and liquid phases are considered incompressible throughout the numerical simulations. Three-dimensional simulation is carried out using the interFoam solver in the open-source platform OpenFOAM[18, 48, 49]. In the solver, the VOF (volume-of-fluid) method[50] is used for the interface tracking of gas-liquid two phases, and the CSF (continuum surface force) model[51] is employed to model the surface tension.

**A. Governing equation**

The governing equations include the continuity equation, the momentum equation, and the phase equation of the incompressible VOF model, which are formulated below:

$$\nabla \cdot \mathbf{U} = 0, \tag{1}$$

$$\frac{\partial \rho \mathbf{U}}{\partial t} + \nabla \cdot (\rho \mathbf{U}\mathbf{U}) = -\nabla p + \nabla \cdot \tau + \rho \mathbf{g} + \mathbf{f}_\sigma, \tag{2}$$

$$\frac{\partial \alpha}{\partial t} + \nabla \cdot (\mathbf{U}\alpha) = 0, \tag{3}$$

where $\mathbf{U}$ denotes the velocity vector of the fluid, $p$ denotes the pressure, and $\mathbf{g}$ represents the acceleration of gravity. $\alpha$ is the volume fraction, with $\alpha = 1$ corresponding to the liquid phase, $\alpha = 0$ to the gas phase, and $0 < \alpha < 1$ representing the interface of gas-liquid. $\mathbf{f}_\sigma = \sigma \kappa \nabla \alpha$ is the surface tension source term based on the CSF model, which is utilized to represent the effect of surface tension. $\sigma$ is the surface tension of the interface, $\kappa = -\nabla \cdot \mathbf{n}$ refers to the interface curvature, and $\mathbf{n} = \nabla \alpha / |\nabla \alpha|$ represents the unit normal vector of the interface. $\tau = \mu (\nabla \mathbf{U} + \nabla \mathbf{U}^T)$ is defined as the viscous stress tensor, and $\mathbf{I}$ corresponds to the unit tensor.

The fluid density and dynamic viscosity are determined based on the weighted averaging of phase fractions:

$$\rho = \alpha \rho_l + (1-\alpha) \rho_g, \tag{4}$$

$$\mu = \alpha \mu_l + (1-\alpha) \mu_g. \tag{5}$$

The subscript "g" in the equation represents the gas phase, and the subscript "l" represents the liquid phase.

For preserving the sharpness of the phase interface, the artificial convection term is added to squeeze the phase fraction near the phase interface, thereby counteracting the phase interface blurring induced by numerical dissipation[52]

$$\frac{\partial \alpha}{\partial t} + \nabla \cdot (\mathbf{U}\alpha) + \nabla \cdot \left[ \alpha(1-\alpha) c_\alpha |\mathbf{U}| \frac{\nabla \alpha}{|\nabla \alpha|} \right] = 0. \tag{6}$$

The third term corresponds to the artificially added convection term, where $c_\alpha$ is the compression coefficient, and its value is 1 in this study.

Finally, the contact angle is introduced into the model as a boundary condition, and its prescribed value is attained by controlling the orientation of the interface normal at the wall boundary[53]. After solving the phase equation and reconstructing the phase interface, a contact angle $\theta_i$ will be formed at the three-phase contact line, differing from the one imposed in the boundary condition. At this time, the normal vector $\mathbf{n}_i$ of the reconstructed phase interface is given by:

$$\mathbf{n}_i = \mathbf{n}_w \cos \theta_i + \mathbf{n}_t \sin \theta_i, \tag{7}$$

where $\mathbf{n}_w$ corresponds to the unit normal vectors of the vertical wall, while $\mathbf{n}_t$ corresponds to that of the parallel wall. The phase interface normal vector $\mathbf{n}_{eq}$ corresponding to the contact angle $\theta_{eq}$ set in the boundary condition can be expressed as:

$$\mathbf{n}_{eq} = \mathbf{n}_w \cos \theta_{eq} + \mathbf{n}_t \sin \theta_{eq}. \tag{8}$$

From Eqs. (7) and (8), we can obtain

$$\mathbf{n}_{eq} = a\mathbf{n}_w + b\mathbf{n}_i \quad (9)$$

$$a = \frac{\cos\theta_{eq} - \cos\theta_i \cos(\theta_i - \theta_{eq})}{1 - \cos^2\theta_i}, \quad (10)$$

$$b = \frac{\cos(\theta_i - \theta_{eq}) - \cos\theta_i \cos\theta_{eq}}{1 - \cos^2\theta_i}. \quad (11)$$

Based on the above-mentioned equations, the modified gas-liquid interface can be calculated by $\mathbf{n}_w$, $\mathbf{n}_i$, $\theta_i$, and $\theta_{eq}$, so as to realize the constraint condition of the contact angle. In this study, $\theta_{eq}$ is set as a constant, i.e., the static contact angle model.

The finite volume method (FVM) is utilized for the solution of the governing equations. The second-order upwind method is applied to discretize the momentum equation and the continuity equation. Temporal discretization is carried out with the Euler method, whereas the Gaussian linear scheme is employed to discretize the gradient term. The coupling between pressure and velocity is achieved using the PIMPLE (pressure implicit method for pressure linked equations) algorithm, which combines the SIMPLE (semi-implicit method for pressure linked equations) and PISO (pressure implicit with splitting of operators) methods. The algorithm has high stability and efficiency[54]. The MULES (multi-dimensional universal limiter with explicit solution)[55] scheme, based on the FCT (flux correct transport)[56] algorithm, is employed to maintain the boundedness of the phase fraction in the phase equation. The simulations are performed using an adaptive time step, and the Courant number is constrained to a maximum value of 0.2.

## B. Physical model

A schematic illustration of the droplet impact on the cylinder is provided in Fig. 1(a). A droplet of diameter $D_d$ impacts the cylinder, which has a diameter of $D_c$, from directly above with an initial velocity of $v_i$. After contacting the cylinder, spreading of the droplet occurs in both the azimuthal direction [Fig. 1(b)] and the axial direction [Fig. 1(c)]. In this study, the azimuthal spreading angle of the droplet is represented by $\omega$, the azimuthal spreading length is denoted as $L_{az}$ [corresponding to the red line in Fig. 1(b)], and the axial spreading length is represented as $L_{ax}$. The diameter ratio $D^* = D_c/D_d$ is expressed as the ratio of the cylinder diameter to the droplet diameter, and the Ohnesorge number $Oh = \mu_l / \sqrt{\rho_l D_d \sigma}$ is employed to represent the relative magnitude of viscous forces. In all simulation cases, a gravitational acceleration of 9.8 m/s² is applied in the simulations, the droplet diameter is $D_d$ = 2.5 mm, the cylinder diameter is $D_c$ = 10 mm. The densities of the liquid and gas phases are taken as $\rho_l$ = 997 kg/m³ and $\rho_g$ = 1.184 kg/m³, respectively, with a surface tension coefficient of $\sigma$ = 0.072 N/m and a gas dynamic viscosity of $\mu_g$ = 0.0184 mPa s.

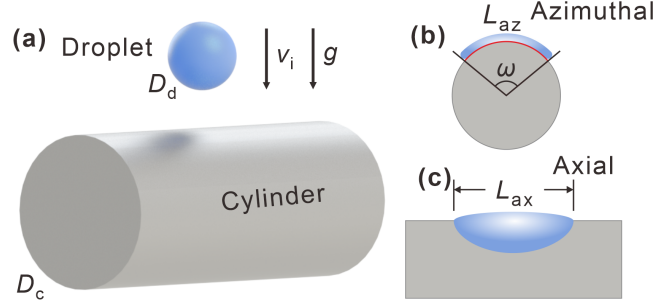

FIG. 1. Physical model of droplet impact on a cylinder: (a) schematic illustration of the droplet impacting the cylinder; (b) azimuthal spreading process, showing the azimuthal spreading angle $\omega$ and azimuthal spreading length $L_{az}$; and (c) axial spreading process, showing the axial spreading length $L_{ax}$.

## C. Mesh independence study

In this study, hexahedral structured meshes are used for the numerical simulations. To accurately capture the movement of droplets near the wall, the mesh is refined near the wall. The cylindrical surface is specified as a no-slip boundary, whereas pressure outlet boundaries are assigned to all other surfaces. The mesh independence is verified by comparing the change in the azimuthal spreading angle [as depicted in Fig. 2(a)] and the axial spreading length [as depicted in Fig. 2(b)] with time under different grid sizes. The results indicate that when the total number of cells reaches $6.4\times10^6$, further mesh refinement leads to no significant changes in $\omega$ and $L_{az}$. Therefore, to balance computational accuracy and efficiency, a mesh with $6.4\times10^6$ cells is adopted in this study.

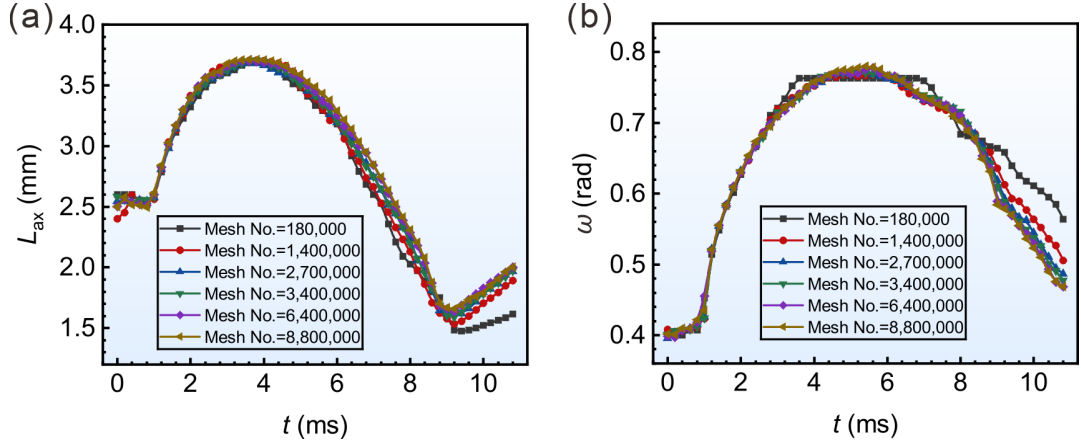

FIG. 2. Verification of mesh independence: (a) temporal variation of the axial spreading length $L_{az}$ under various mesh sizes; and (b) temporal variation of the azimuthal spreading angle $\omega$ under various mesh sizes. Here, $v_i = 0.5$ m/s, $\mu_l = 0.898$ mPa·s, $\theta_{eq} = 160°$, $We = 8.65$, $Oh = 0.00213$, and $D^* = 4.0$.

## D. Model validation

The reliability of the numerical model was assessed by comparing the simulation results with the experimental data reported by *Liu et al.*[25]. Simulation conditions are established following the experimental setup of droplet impact on a superhydrophobic horizontal cylinder. The specific parameters are as follows: cylinder diameter $D_c = 8.0$ mm, droplet diameter $D_d = 2.9$ mm, impact velocity $v_i = 0.63$ m/s, equilibrium contact angle $\theta_{eq} = 163.4°$, and Weber number $We = 7.9$. As illustrated in Figs. 3(a) and 3(b), the temporal evolution of the droplet morphology from numerical simulations is in good agreement with the experiment. Figure 3(c) illustrates the evolution of the contact line length in both the axial and azimuthal directions throughout the impact process. According to the simulation results, the maximum axial spreading length is 4.61 mm at 4 ms, and the contact time is 11.4 ms. These values differ from the experimental data (4.45 mm and 11.2 ms, respectively) by only 3.6% and 1.8%. Hence, the numerical results obtained in this study are in close agreement with the experimental measurements. The numerical model is reliable and can be used for the subsequent simulations.

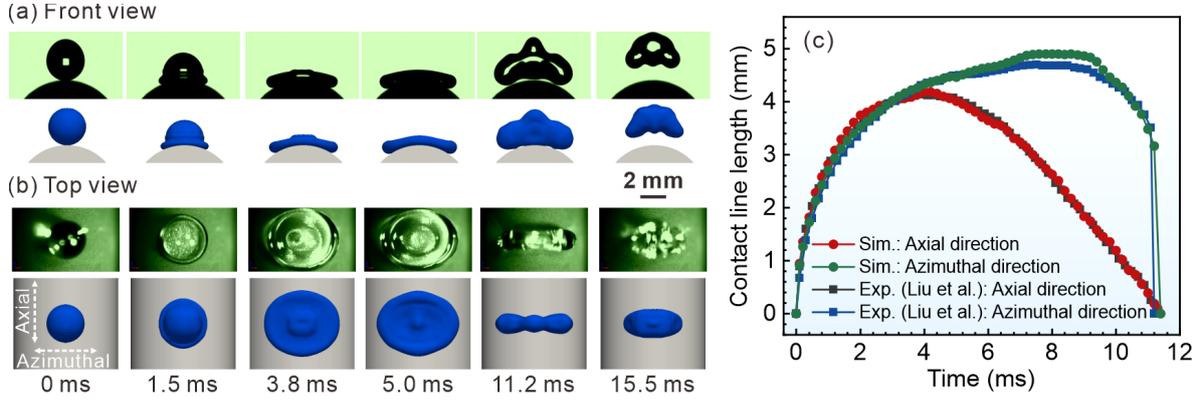

FIG. 3. Comparison between simulated and experimental results showing the droplet morphology and the evolution of contact line length during impact: (a) front view of the droplet morphology; (b) top view of the droplet morphology; and (c) contact line length variation along the azimuthal and axial directions.

The temporal behavior of the impact force during droplet impact on a cylinder is investigated in this study, and the accuracy of the force calculation method is confirmed. The impact force generated by the droplet upon the surface can be decomposed into two components: the pressure force acting on the contact area and the surface tension force applied along the three-phase contact line[38]. As the impact process investigated here is symmetrical, the resultant impact force is directed along the gravitational direction, and its magnitude can be expressed as follows:

$$F = F_h - F_t, \tag{12}$$

where $F$ denotes the resultant impact force, and $F_h$ represents the force generated by the pressure exerted by the droplet on the wall, obtained by integrating the pressure distribution over the liquid-solid interface:

$$F_h = \int_{A_{sl}} \left[ (p - p_0)(\mathbf{n}_w \cdot \mathbf{z}) \right] dA_{sl}. \tag{13}$$

$F_t$ represents the surface tension acting along the three-phase contact line, which can be calculated by integrating along the contact line:

$$F_t = \int_{L_{cl}} \sigma \sin\theta (\mathbf{n}_w \cdot \mathbf{z}) dL_{cl}, \tag{14}$$

where $A_{sl}$ represents the solid-liquid interface area, $p_0$ denotes the atmospheric pressure, $p$ is the pressure exerted on the solid wall, $\mathbf{z}$ indicates the unit vector in the direction of gravity, and $L_{cl}$ stands for the length of the three-phase contact line.

Since experimental data on the impact force of droplets impinging on cylindrical surfaces are unavailable, the present study employs the experimental data of flat walls reported by Zhang et al.[38] for validation, as presented in Fig. 4. In their experiment, the normal impact force between the drop and the wall was measured by collecting the charge generated by the highly sensitive piezoelectric transducer, which is instantly converted into a voltage through an amplifier, and converted to digital signals using a data acquisition system. Ultimately, the measured signal was converted through the calibration coefficient of the force transducer. The comparison reveals that the simulated impact forces are in excellent agreement with the experimental measurements for both hydrophilic and hydrophobic surfaces, indicating that the proposed method for impact force calculation is reliable.

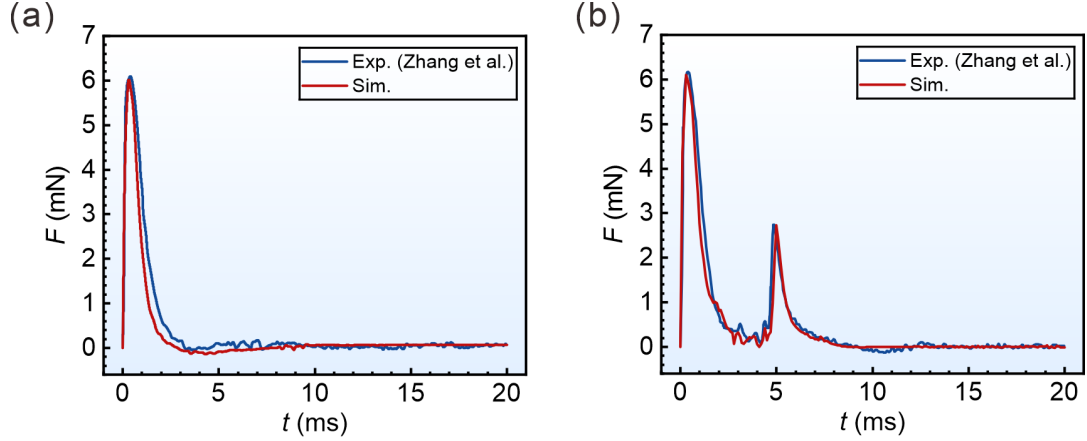

FIG. 4. Temporal variation of the impact force on flat surfaces: (a) hydrophilic surface ($\theta_{eq}$ = 30°); (b) superhydrophobic surface ($\theta_{eq}$ =161°). Here, $D_d$ = 2.05 mm, $v_i$ = 1.33 m/s, $\rho_l$ = 1000 kg/m³, $\sigma$ = 0.073 N/m, $\mu_l$ = 1.00 mPa·s, $We$ = 49.7, and $Oh$ = 0.00259.

## III. Results and discussion

### A. Typical process of impact on hydrophilic and hydrophobic cylinders

Our numerical results indicate that droplet impact on hydrophilic and hydrophobic cylinders exhibits markedly distinct interfacial evolution patterns and impact force characteristics. Droplet impact on a hydrophilic wall results in deposition, during which the impact force exhibits a single peak. In contrast, droplet impact on a hydrophobic wall is characterized by rebound behavior accompanied by two peaks in the impact force. Accordingly, we first describe the representative processes of droplet impact under typical conditions on a hydrophilic wall (with contact angle $\theta_{eq}$ = 60°) and on a hydrophobic wall (with contact angle $\theta_{eq}$ = 150°).

#### 1. Droplet impact on a hydrophilic cylinder

The dynamic evolution of a droplet impacting a hydrophilic cylinder under the deposition mode is shown in Fig. 5. As shown in Fig. 5(a) (Multimedia view), the process can be classified into three successive stages. The droplet spreads synchronously in the axial and azimuthal directions during the first stage, reaching its maximum spreading length in the axial direction at 5.3 ms. The droplet starts to retreat in the axial direction during the second stage, while unlike droplet impact on a flat surface, it keeps spreading in the azimuthal direction until it reaches its maximum azimuthal spreading length at 8.3 ms. This is because the contact area formed when the droplet impacts the cylinder is elliptical, with more momentum being transferred along the major axis of the ellipse (i.e., the azimuthal direction). As a consequence, the maximum spreading lengths in the azimuthal and axial directions are not achieved simultaneously. In the third stage, both the azimuthal and axial spreading lengths begin to retract at the same time, followed by oscillations in both directions until the droplet eventually stabilizes, forming a clam-shell shaped droplet, consistent with the static droplet shape on hydrophilic cylinders[57].

To provide a quantitative characterization of droplet kinematic evolution during impact on cylinders, the dimensionless spreading area $S^* = S/\left(\pi(D_d/2)^2\right)$, dimensionless axial spreading length $L_{ax}^* = L_{ax}/D_d$, and dimensionless azimuthal spreading angle $\omega^* = \omega/2\pi$ are defined, as illustrated in Fig. 1(b) and 1(c). The instants corresponding to their respective maxima are denoted as $t_{area,max}$, $t_{az,max}$, and $t_{ax,max}$, respectively. For the hydrophilic surface, the droplet undergoes continuous oscillations on the cylinder before reaching a steady state; therefore, data within the first 100 ms are considered in the analysis. As shown in Fig. 5(b), we can find that $t_{ax,max} < t_{area,max} < t_{az,max}$. Both $S^*$ and $L_{ax}^*$ increase initially, then decrease, and finally oscillate until they become stable. Meanwhile, $\omega^*$ oscillates slightly at the beginning of the third stage. This behavior arises from the anisotropy of the cylindrical geometry, which induces a transfer of momentum and mass from the axial to the azimuthal direction as the droplet simultaneously retracts in both directions. Consequently, the reduction of $\omega^*$ is delayed compared with $L_{ax}^*$, and $\omega^*$ may even stagnate or slightly increase[58].

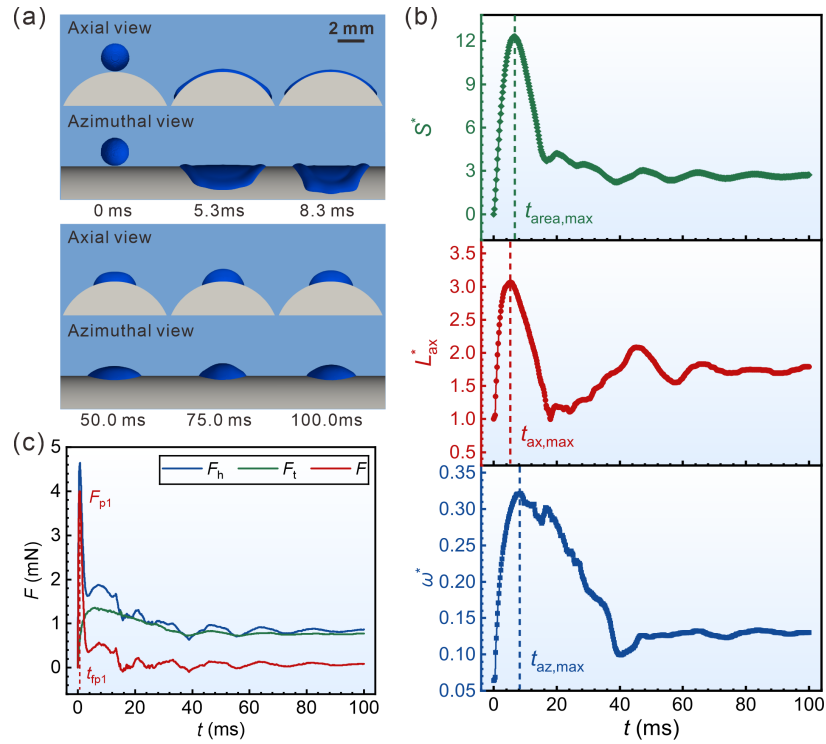

FIG. 5. Dynamic evolution of a droplet impacting a hydrophilic cylinder ($\theta_{eq} = 60°$): (a) evolution of the droplet morphology over time; (b) evolution of the kinematic parameters, including the dimensionless spreading area $S^*$, dimensionless axial spreading length $L_{ax}^*$, and dimensionless azimuthal spreading angle $\omega^*$ over time; (c) temporal evolution of the impact force. Here, $v_i = 1.0$ m/s, $\mu_l = 0.898$ mPa·s, $We = 34.6$, $Oh = 0.00213$, and $D^* = 4.0$. Multimedia available online.

Figure. 5(c) depicts the time evolution of impact force when droplets impact a hydrophilic cylinder. When the droplet makes contact with the wall, the force quickly reaches its peak value ($F_{p1}$ = 4.02 mN) within a very short time ($t_{fp1}$ = 0.47 ms). As the droplet spreads over the surface, the impact force $F$ decreases. This reduction can be attributed to two factors. First, the velocity direction of the droplet changes from vertically downward to tangential along the cylinder, leading to a decrease in $F_h$. Second, $F_t$ gradually increases during spreading. The combined effect of these two components results in the overall decline of $F$. Subsequently, as the droplet oscillates on the wall, $F$ also exhibits periodic oscillations before eventually approaching 0. Similar to droplet impact on hydrophilic flat surfaces[32-35, 46], only a single significant peak in impact force is observed during impact on the hydrophilic cylindrical surface.

**2. Droplet impact on hydrophobic cylinder**

The dynamic behavior of the droplet during impact and rebound on a hydrophobic cylinder is presented in Fig. 6(a) (Multimedia view). Similar to the impact on the hydrophilic cylinder, the overall process can also be classified into three successive stages. The droplet reaches its maximum axial and azimuthal spreading extents at 2.9 and 4.1 ms, respectively. However, in the third stage, no obvious oscillations are observed. The conversion of kinetic energy into surface energy is sufficient for the droplet to rebound. During the retraction stage, the droplet shape on the cylinder can be approximated as an ellipse with semi-major and semi-minor axes denoted by $a$ and $b$, respectively. An approximate expression for the surface energy $E_s$ can be written as:

$$E_s \approx \pi a b \sigma \left(1 - \cos\theta_{eq}\right). \tag{15}$$

The retraction forces along the axial and azimuthal directions are represented by $F_{ax}$ and $F_{az}$, can be expressed as follows:[25, 59]

$$F_{ax} = \frac{\partial E_s}{\partial a} \approx \pi b \sigma \left(1 - \cos\theta_{eq}\right), \tag{16}$$

$$F_{az} = \frac{\partial E_s}{\partial b} \approx \pi a \sigma \left(1 - \cos\theta_{eq}\right). \tag{17}$$

Therefore, the axial retraction of the liquid film is faster. Once the axial liquid film contracts to the central region, the droplet begins to move upward, and it is lifted off at 9.9 ms, forming a wing-like shape[26, 60] and detaching from the cylinder. For the hydrophobic surface, the data are recorded until the moment the droplet just detaches from the wall, as illustrated in Fig. 6(b). Similar to the hydrophilic surface $t_{ax,max} < t_{area,max} < t_{az,max}$. The dimensionless spreading area $S^*$ increases at first and then decreases continuously to 0 until rebound. In the third stage, the dimensionless axial spreading length $L^*_{ax}$ does not oscillate as on the hydrophilic surface, but instead shows a slight increase before droplet rebound. This is because, during retraction, the axial liquid film retracts faster and first reaches the center, then moves upward and gathers, slightly increasing the axial spreading length. Similar to the hydrophilic cylinder, at the early time of the third stage, momentum and mass transfer from the axial to the azimuthal direction causes the dimensionless azimuthal spreading angle $\omega^*$ to decrease slowly.

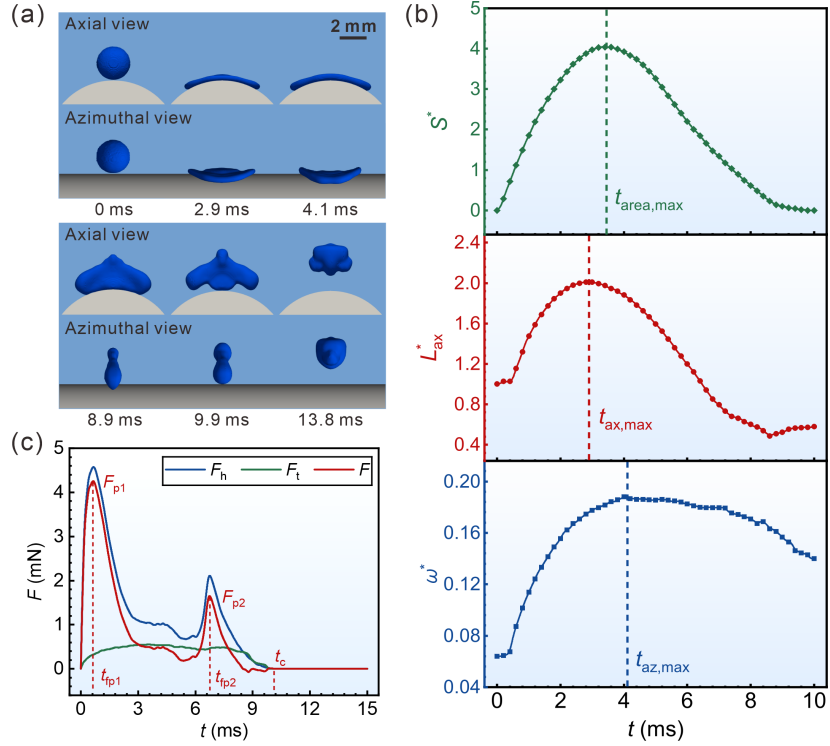

FIG. 6. Dynamic evolution of a droplet impacting a hydrophobic cylinder ($\theta_{eq} = 150°$): (a) temporal evolution of the droplet morphology; (b) evolution of the kinematic parameters, including the dimensionless spreading area $S^*$, dimensionless axial spreading length $L^*_{ax}$, and dimensionless azimuthal spreading angle $\omega^*$; and (c) evolution of the impact force. Here, $v_i = 1.0$ m/s, $\mu_l = 0.898$ mPa·s, $We = 34.6$, $Oh = 0.00213$, and $D^* = 4.0$. Multimedia available online.

The variation of impact force over time during droplet impingement on a hydrophobic cylinder is illustrated in Fig. 6(c). Similar to the hydrophilic case, immediately after impact, the impact force rises sharply and reaches its first peak ($F_{p1} = 4.27$ mN) within a very short time ($t_{fp1} = 0.62$ ms). As the liquid film spreads along the surface, the impact force $F$ begins to decrease. During the subsequent retraction stage, the liquid film contracts toward the center and generates an upward jet, leading to the appearance of a second peak $F_{p2} = 1.68$ mN at $t_{fp2} = 6.7$ ms. This second peak is typically referred to as the jump-off force[37]. With the droplet gradually detaching from the cylindrical surface, $F$ gradually decreases and approaches 0 when complete detachment occurs at $t_c = 9.9$ ms. Unlike the impact on the hydrophilic cylinder, where only a single force peak is observed, two distinct peaks appear for the hydrophobic surface. The first peak results from the rapid transfer of the droplet's initial momentum upon impact, whereas the second peak arises from the reaction force induced by the jet formed during the retraction process.

**B. Effect of surface wettability**

As discussed in Sec. III.A, the wettability of the wall plays a crucial role in the impact behavior of the droplet. In this section, the effects of wettability on the kinematic and dynamic responses of droplet impact on a cylinder are investigated by changing the equilibrium contact angle $\theta_{eq}$. Figure 7(a) presents the change in spreading area over time during droplet impact on surfaces of different wettabilities. When $\theta_{eq}$ exceeds 120°, the droplet rebounds after impacting the cylinder, whereas for contact angles below 120°, only deposition occurs. The findings demonstrate that both the deformation and rebound behaviors of the droplet are strongly dependent on surface wettability. To account for variations in droplet size, impact velocity, and surface tension, and to ensure comparability under different conditions, the time is nondimensionalized with the inertial-capillary timescale:

$$t_\sigma = \left(\rho_l R_d^3 / \sigma\right)^{1/2}, \tag{18}$$

where $R_d$ denotes the droplet radius. As depicted in Fig. 7(b), both the dimensionless time to reach the maximum spreading area $t_{area,max} / t_\sigma$ and the dimensionless maximum spreading area $S^*$ decrease with the increase in $\theta_{eq}$. The dimensionless maximum azimuthal spreading angle $\omega^*$ and dimensionless axial spreading length $L_{ax}^*$ gradually decrease as $\theta_{eq}$ increases, as demonstrated in Fig. 7(c). According to Eqs. (16) and (17), during the spreading process, the resistive forces acting in both azimuthal and axial directions increase with the contact angle. Consequently, for droplets with identical initial kinetic energy, $\omega^*$ and $L_{ax}^*$ both decrease as the contact angle increases. Because mass and momentum are transferred from the axial to the azimuthal direction, the dimensionless time to reach the maximum axial length $t_{ax,max} / t_\sigma$ is shorter than that for the maximum azimuthal spreading angle $t_{az,max} / t_\sigma$, and both decrease with increasing contact angle, as shown in Fig. 7(d). To provide a quantitative description of the droplet spreading asymmetry, an asymmetry coefficient is defined as $L^* = L_{az,max} / L_{ax,max}$, where $L_{az,max}$ and $L_{ax,max}$ correspond to the maximum spreading lengths of the liquid film in the azimuthal and axial directions, respectively. A larger $L^*$ indicates that more mass and momentum are transferred from the axial to the azimuthal direction. A reduction in wettability leads to a decrease in $L^*$, as illustrated in Fig. 7(e). Once the rebound occurs, the effect of surface wettability on $L^*$ becomes negligible, and it stabilizes at approximately 1.17.

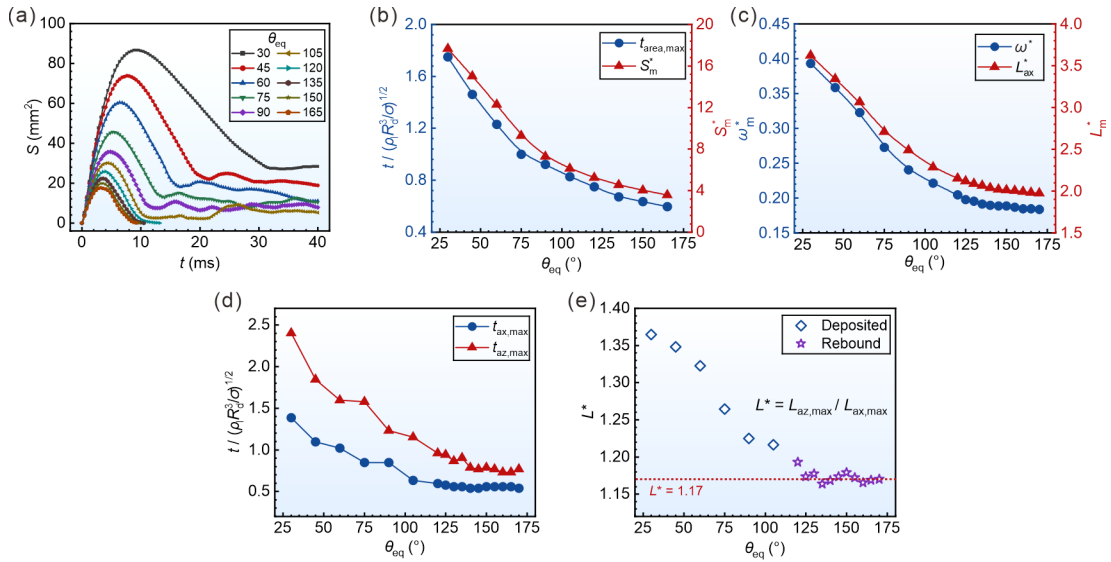

FIG. 7. Effect of surface wettability on the kinematic characteristics of droplet impact on a cylinder: (a) evolution of the spreading area over time under different contact angles; (b) variation of $S_m^*$ and $t_{area,max} / t_\sigma$ with surface wettability; (c) variation of $\omega^*$ and $L_{ax}^*$ with surface wettability; (d) variation of $t_{ax,max} / t_\sigma$ and $t_{az,max} / t_\sigma$ with surface

wettability; and (e) variation of the asymmetry coefficient $L^*$ with surface wettability. Here, $v_i$ = 1.0 m/s, $\mu_l$ = 0.898 mPa·s, $We$ = 34.6, $Oh$ = 0.00213, and $D^*$ = 4.0.

The wettability of the cylindrical surface influences not only the asymmetrical deformation exhibited by the droplet in the spreading and rebound stages but also the impact force generated upon impact. Figures 8(a) and 8(b) display the variation of the transient impact force with contact angle in the deposition and rebound regimes, respectively. For the deposition regime, the numerical simulations are carried out to 40 ms to ensure that the droplet reaches the oscillation stage, while for the rebound regime, the simulations are carried out to 15 ms to capture the entire rebound process. The results indicate that, for droplets impacting both hydrophilic and hydrophobic cylindrical surfaces, the impact force curves on both sides of the first peak almost overlap, exhibiting similar evolution. This implies that the impact force is only weakly affected by the contact angle during the inertia-dominated initial stage. In the rebound regime, a second peak appears in the impact force curve, which is notably sensitive to surface wettability. During the retraction stage, wettability affects the magnitude of the capillary retraction force, thereby influencing the retraction velocity, the timing of the second peak, and the contact time. In addition, the magnitude of the impact force during droplet rebound is strongly affected by wettability. Figure 8(c) displays the change in the dimensionless time $t_{fp1} / t_\sigma$ corresponding to the first peak with surface wettability. The results show that $t_{fp1} / t_\sigma$ increases slightly with increasing contact angle and approaches a constant value under hydrophobic conditions, where $t_{fp1} / t_\sigma \approx 0.12$. For the rebound regime, both the dimensionless contact time $t_c / t_\sigma$ and the dimensionless time of the second peak $t_{fp2} / t_\sigma$ decrease as the contact angle increases. To better interpret the impact force behavior, the rebound process of a droplet impacting a solid wall is analogous to the simple harmonic oscillation of a water spring system[61]. In such a system, the oscillation period depends on both the oscillator mass and its stiffness coefficient, which are analogous to the droplet mass and surface tension coefficient during impact, respectively. Therefore, the characteristic time of the impact process can be expressed as:

$$t \sim t_\sigma = \left( \rho_l R_d^3 / \sigma \right)^{1/2}. \tag{19}$$

Based on Hooke's law $F = ks$, considering the influence of the wall contact angle on the droplet impact process, the stiffness coefficient $k$ of the water spring system (which corresponds to the surface tension coefficient in the impact process) can be modified as $\sigma \left(1 - \cos\theta_{eq}\right)$[62] based on Eqs. (16) and (17). Therefore, when the droplet impacts the hydrophobic cylinder and rebounds, the characteristic time is modified as:

$$t \sim t'_\sigma = \left[ \frac{\rho_l R_d^3}{\sigma \left(1 - \cos\theta_{eq}\right)} \right]^{1/2} = t_\sigma \left(1 - \cos\theta_{eq}\right)^{-1/2}, \tag{20}$$

$$t/t_\sigma \sim \left(1 - \cos\theta_{eq}\right)^{-1/2}. \tag{21}$$

As demonstrated in Fig. 8(d), both the dimensionless time to reach the second peak $t_{f2} / t_\sigma$ and the dimensionless contact time $t_c / t_\sigma$, satisfy the above relationship.

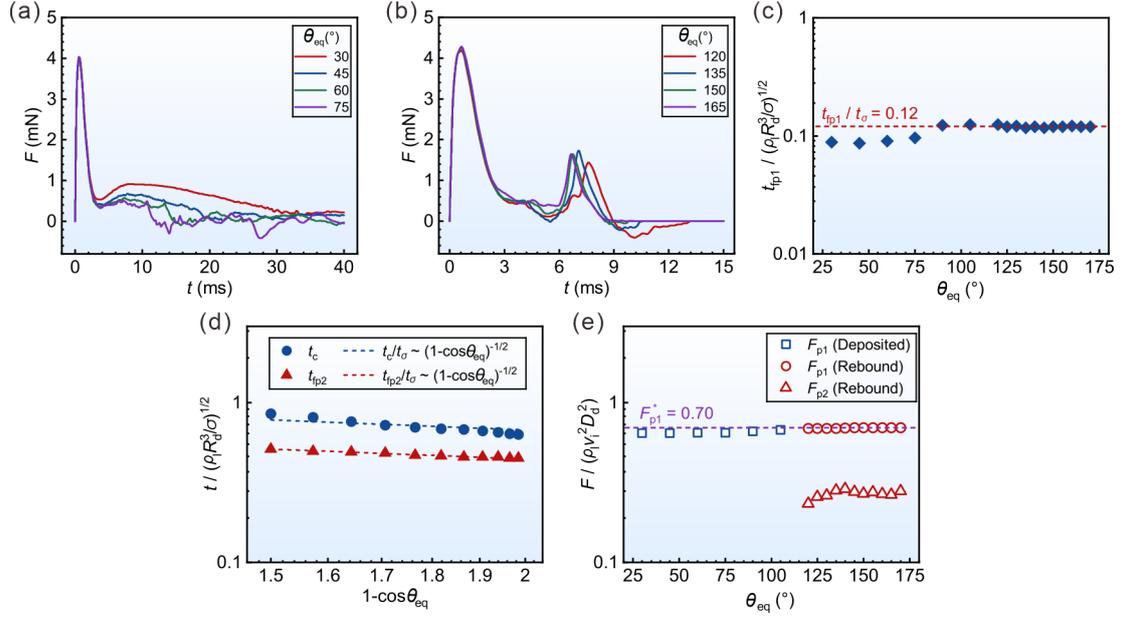

FIG. 8. Effect of wettability on the dynamic characteristics of droplet impact on a cylinder: (a) development of the impact force under various contact angles for deposition regime; (b) evolution of the impact force under various contact angles for rebound regime; (c) dependence of the dimensionless time $t_{fp1}/t_\sigma$ on contact angle; (d) dependence of the dimensionless times $t_{fp2}/t_\sigma$ and $t_c/t_\sigma$ on contact angle; and (e) dependence of the dimensionless peak impact forces on contact angle. Here, $v_i = 1.0$ m/s, $\mu_l = 0.898$ mPa·s, $We = 34.6$, $Oh = 0.00213$, and $D^* = 4.0$.

The impact force is directly related to the droplet's inertial force during the impact process. Therefore, the peak impact forces $F_{p1}$ and $F_{p2}$ are nondimensionalized by $\rho_l v_i^2 D_d^2$ to obtain $F_{p1}^* = F_{p1}/\rho_l v_i^2 D_d^2$ and $F_{p2}^* = F_{p2}/\rho_l v_i^2 D_d^2$, as depicted in Fig. 8(e). It is observed that wettability has a negligible effect on the first nondimensional peak impact force $F_{p1}^*$. It increases slightly with increasing contact angle and approaches a constant value of 0.70 under rebound conditions. This is because, in the initial stage of impact, the variation exhibited by impact force is mainly governed by the change in vertical momentum, where inertial effects dominate and surface tension plays a minor role. In addition, it has been reported that in the case of impact on a flat wall, $F_{p1}^* = 0.81$[36], which is greater than the value obtained for a cylindrical surface. This difference arises because when the surface has curvature, the droplet spreads along the azimuthal direction after impact, retaining part of its vertical momentum. As a result, the overall impact force is smaller than that on a flat wall. The second nondimensional peak impact force $F_{p2}^*$ increases slightly with the contact angle and stabilizes at approximately 0.27 when the surface becomes superhydrophobic ($\theta_{eq} > 150°$).

## C. Effect of Weber number

The Weber number $We$ quantifies the balance between inertial and surface tension forces in the droplet impact process, making it a crucial dimensionless parameter in characterizing impact behavior. In this section, $We$ is varied from 9 to 138 by adjusting the droplet impact velocity, whereas the other parameters are maintained constant. To capture different impact regimes, two types of cylinders with equilibrium contact angles of 60° and 150° are considered. For the deposition regime, Fig. 9(a) presents the temporal evolution of the spreading area under different $We$ conditions. The spreading area undergoes three distinct stages: initial increase, subsequent decrease, and final oscillation until reaching equilibrium. In contrast, for the rebound regime, the spreading area only experiences two stages, expansion and subsequent contraction to 0, as shown in Fig. 9(b). In both regimes, an increase in $We$ leads to a larger maximum spreading area $S$ and a shorter time to reach it. This behavior occurs because a higher $We$ corresponds to greater initial kinetic energy, which upon impact is converted into surface energy, expanding the maximum spreading area. Moreover, stronger inertial effects at higher $We$ promote faster energy conversion, thus shortening the time required to reach the maximum spreading area. As presented in Fig. 9(c), the dimensionless maximum spreading area $S_m^*$ exhibits a power-law relationship with $We$, which can be obtained through fitting as:

$$S_m^* \sim \begin{cases} We^{0.42} & \theta_{eq} = 60° \\ We^{0.69} & \theta_{eq} = 150° \end{cases}. \tag{22}$$

The time variable was normalized using the inertial timescale $t_\rho = D_d/v_i$. As depicted in Fig. 9(d), the dimensionless time to reach the maximum spreading area $t_{area,max}/t_\rho$ increases with increasing $We$ and also follows a power-law relationship. The fitted relationship can be expressed as:

$$t_{area,max}/t_\rho \sim \begin{cases} We^{0.33} & \theta_{eq} = 60° \\ We^{0.44} & \theta_{eq} = 150° \end{cases}. \tag{23}$$

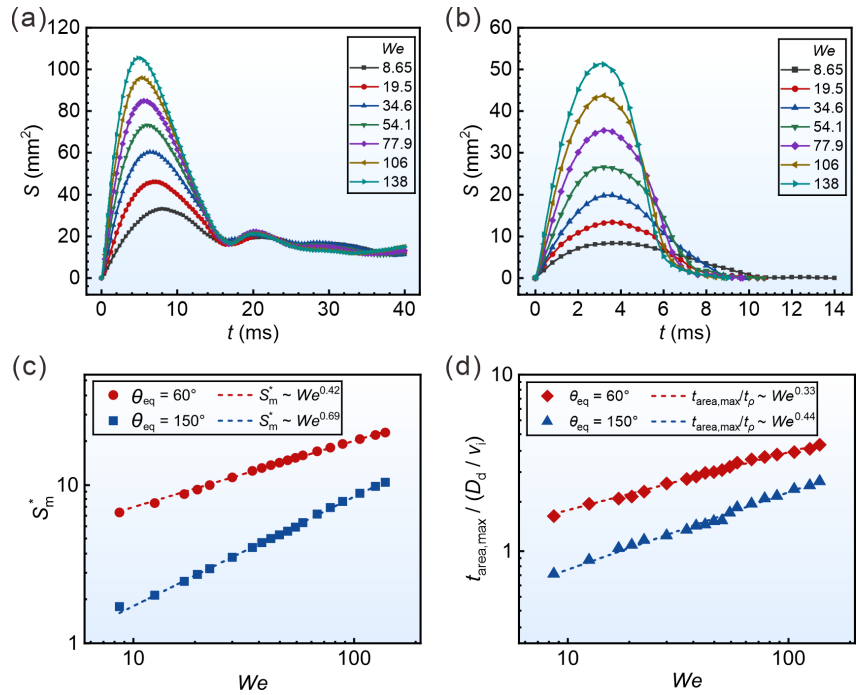

FIG. 9. Effects of $We$ on the spreading area of droplet impact on a cylinder: (a) evolution of spreading area in the deposition regime ($\theta_{eq}$ = 60°); (b) evolution of spreading area in the rebound regime ($\theta_{eq}$ = 150°); (c) variation of dimensionless maximum spreading area $S_m^*$ with $We$; and (d) dependence of the dimensionless time to reach maximum spreading area $t_{area,max} / t_\rho$ on $We$. Here, $8 \leq We \leq 138$, $\mu_l = 0.898$ mPa·s, $Oh = 0.00213$, and $D^* = 4.0$.

The droplet spreading length and angle are also influenced by $We$. Figures 10(a) and 10(b) present the variations of the dimensionless maximum azimuthal spreading angle and the axial spreading length with $We$, respectively. Both $\omega_m^*$ and $L_m^*$ increase with increasing $We$, as a higher $We$ corresponds to greater droplet kinetic energy and relatively weaker surface tension effects, making the droplet easier to spread. Fitting the data of the maximum azimuthal spreading angle and axial spreading length yields, the following can be obtained:

$$\omega_m^* \sim \begin{cases} We^{0.22} \\ We^{0.36} \end{cases}, \quad L_m^* \sim \begin{cases} We^{0.19} & \theta_{eq} = 60° \\ We^{0.27} & \theta_{eq} = 150° \end{cases}, \tag{24}$$

From Eq. (24), the relationship between the asymmetry factor $L^*$ and $We$ can be described as:

$$L^* = \frac{L_{az,max}}{L_{ax,max}} \sim \frac{\omega_m^*}{L_m^*} \sim \begin{cases} We^{0.03} & \theta_{eq} = 60° \\ We^{0.09} & \theta_{eq} = 150° \end{cases}. \tag{25}$$

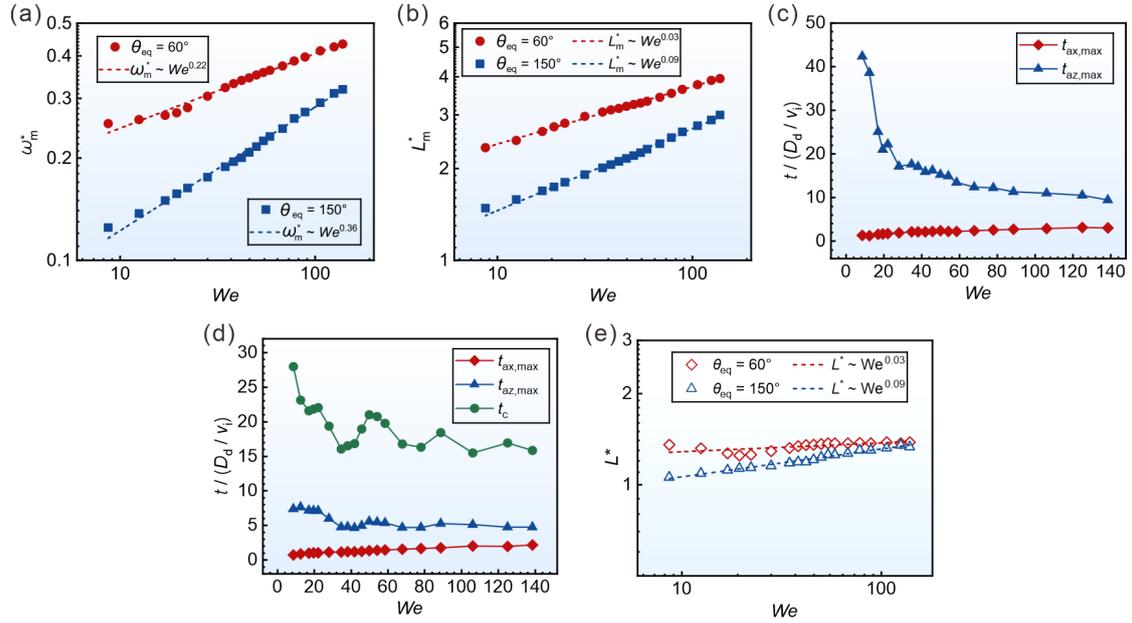

FIG. 10. Influence of $We$ on the azimuthal and axial spreading of droplets impacting a cylinder: (a) dependence of the dimensionless maximum azimuthal spreading angle $\omega_m^*$ on $We$; (b) dependence of the dimensionless maximum axial spreading length $L_m^*$ on $We$; (c) variation of the dimensionless characteristic times $t_{ax,max} / t_\rho$ and $t_{az,max} / t_\rho$ with $We$ under the deposition regime ($\theta_{eq} = 60°$); (d) variation of the dimensionless characteristic times $t_{ax,max} / t_\rho$, $t_{az,max} / t_\rho$ and $t_c / t_\rho$ with $We$ under the rebound regime ($\theta_{eq} = 150°$); and (e) variation of the asymmetry coefficient $L^*$ with $We$. Here, $8 \leq We \leq 138$, $\mu_l = 0.898$ mPa·s, $Oh = 0.00213$, and $D^* = 4.0$.

The impact force exerted on the surface during droplet impact also depends on the $We$ number. Figures 11(a) and 11(b) show the variation of the impact force under different $We$ for the hydrophilic ($\theta_{eq} = 60°$) and hydrophobic ($\theta_{eq} = 150°$) surfaces over time, respectively. It can be observed that $We$ does not alter the overall trend of impact force variation, but significantly influence the peak magnitude and the timing of its occurrence. As $We$ increases, the droplet possesses higher initial kinetic energy and momentum, causing more rapid deformation and momentum transfer during impact, which enhances the magnitude of the first peak impact force and brings about an earlier occurrence of this peak. The magnitude of the first peak impact force $F_{p1}$, exhibits a linear relationship with $We$, as displayed in Fig. 11(c). Figure 11(d) presents the variation of the dimensionless characteristic times with $We$, where the dimensionless time to reach the first peak impact force $t_{fp1}/t_\rho$ remains around 0.23, whereas the dimensionless time to reach the second peak impact force $t_{fp2}/t_\rho \sim We^{0.35}$. The change in the first dimensionless peak impact force $F_{p1}^*$ with $We$ is shown in Fig. 11(e). For large $We$, both the deposition and rebound regimes approach a constant value of approximately 0.63. At lower $We$, it deviates from this value, similar to observations in droplet impacts on flat surfaces[32-34, 36]. The reason for this phenomenon is that when $We$ is large, the inertial force dominates, and the influence of surface tension relative to the inertial force can be ignored. As $We$ decreases, the influence of surface tension strengthens and cannot be ignored, leading to the first peak impact force $F_{p1}$ and $F_{p1}^*$ being expressed as:[36]

$$F_{p1} = \beta_1 \left( \rho_l v_i^2 D_d^2 \right) + \beta_2 \left( \sigma D_d \right), \tag{26}$$

$$F_{p1}^* = \beta_1 + \beta_2 We^{-1}, \tag{27}$$

where $\beta_1$ and $\beta_2$ are constant coefficients reflecting the relative contributions of inertial and surface tension forces. By fitting the data, the following scaling can be obtained:

$$F_{p1}^* = \frac{F_{p1}}{\rho_l v_i^2 D_d^2} \approx \begin{cases} 0.63 + 0.67 We^{-1} & \theta_{eq} = 60° \\ 0.63 + 2.15 We^{-1} & \theta_{eq} = 150° \end{cases}. \tag{28}$$

As illustrated by the dashed line in Fig. 11(e), this function effectively captures the variation of $F_{p1}^*$ with $We$. Figure 11(f) depicts the dependence of the second dimensionless peak impact force $F_{p2}^*$ on $We$. It can be seen that when $We$ is relatively small, $F_{p2}^*$ exhibits significant fluctuations, whereas at larger $We$ it tends to stabilize. This is because at low $We$, the rebound regime transitions from a jug-like rebound to a wing-like rebound[26, 60], resulting in more complex variations in $F_{p2}^*$.

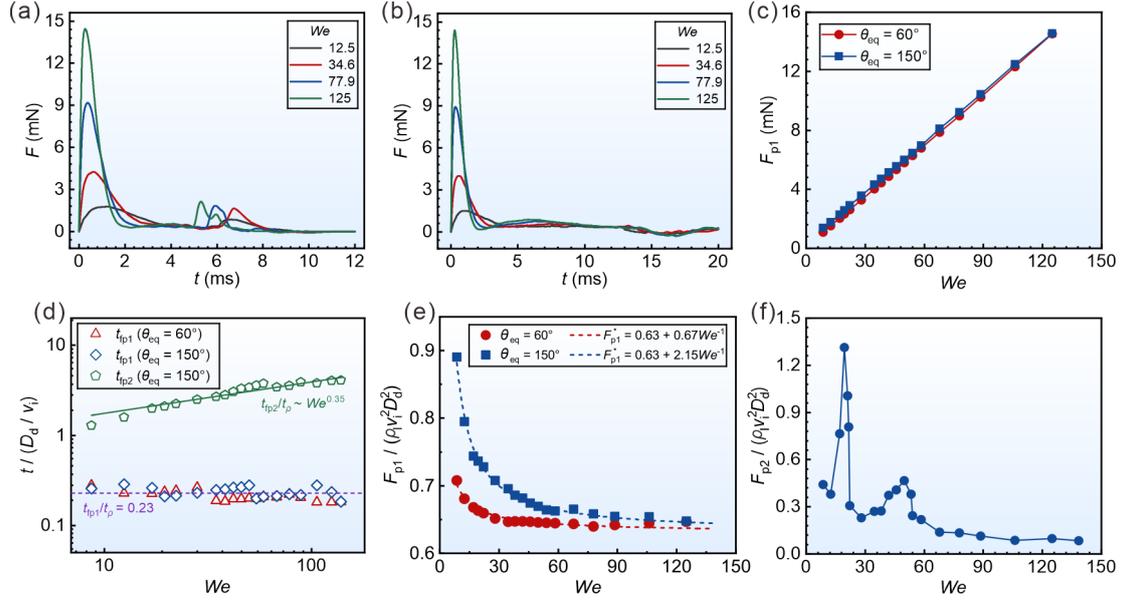

FIG. 11. Effect of *We* on the impact force of droplets impacting a cylinder: (a) evolution of the impact force under deposition regime ($\theta_{eq} = 60°$); (b) evolution of the impact force under rebound regime ($\theta_{eq} = 150°$); (c) variation of the first peak force $F_{p1}$ with *We*; (d) variation of the dimensionless characteristic times $t_{fp1}/t_\rho$ and $t_{fp2}/t_\rho$ with *We*; (e) dependence of the first dimensionless peak force $F_{p1}^*$ on *We*; and (f) dependence of the second dimensionless peak force $F_{p2}^*$ on *We*. Here, $8 \leq We \leq 138$, $\mu_l = 0.898$ mPa·s, $Oh = 0.00213$, and $D^* = 4.0$.

### D. Effect of Ohnesorge number

The Ohnesorge number *Oh* quantifies the relative importance of viscosity in a droplet and has a pronounced effect on its impact dynamics. In this section, *Oh* is varied by altering the viscosity of the liquid. First, the influence of *Oh* on the spreading area is investigated, as shown in Figs. 12(a)-12(c). With increasing *Oh*, the viscous resistance during droplet spreading is enhanced, leading to greater energy dissipation. As a result, a higher *Oh* suppresses droplet spreading and reduces the maximum spreading area. This relationship can be expressed through a fitting as:

$$S_m^* \sim \begin{cases} Oh^{-0.26} & \theta_{eq} = 60° \\ Oh^{-0.10} & \theta_{eq} = 150° \end{cases}. \tag{29}$$

Furthermore, with increasing *Oh*, during the oscillation stage after impacting a hydrophilic surface, the kinetic energy of the droplet decays more rapidly, thereby reducing the oscillation period as well as the amplitude. The dimensionless time to reach the maximum spreading area $t_{area,max}/t_\sigma$ does not change significantly with *Oh*, since the spreading process is still dominated by inertial forces, as shown in Fig. 12(d).

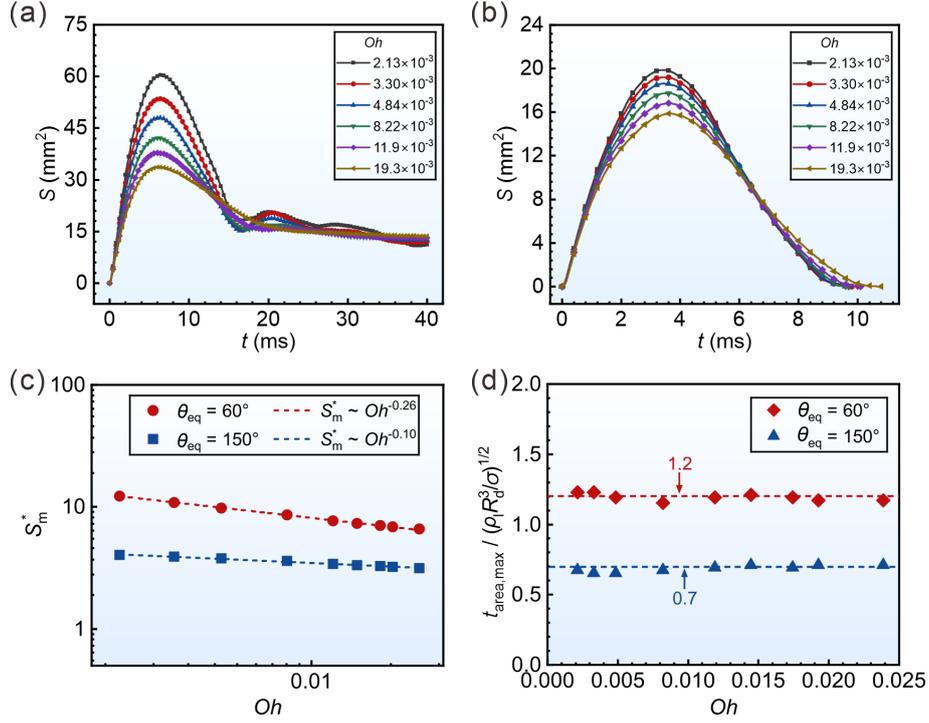

FIG. 12. Effect of $Oh$ on the spreading area of droplet impact on a cylinder: (a) evolution of the spreading area in the deposition regime ($\theta_{eq} = 60°$); (b) evolution of the spreading area in the rebound regime ($\theta_{eq} = 150°$); (c) variation of the dimensionless maximum spreading area $S_m^*$ with $Oh$; and (d) dependence of the dimensionless time to reach the maximum spreading area $t_{area}/t_\sigma$ on $Oh$. Here, $0.00213 \leq Oh \leq 0.0193$, $v_i = 1.0$ m/s, $We = 34.6$, and $D^* = 4.0$.

The spreading length and angle are also affected by the Ohnesorge number during impact. As shown in Figs. 13(a) and 13(b), both the dimensionless maximum spreading angle $\omega_m^*$ and the dimensionless maximum spreading length $L_m^*$ decrease with increasing $Oh$. Through the fitting of the data, we can obtain:

$$\omega_m^* \sim \begin{cases} Oh^{-0.16} \\ Oh^{-0.06} \end{cases}, \quad L_m^* \sim \begin{cases} Oh^{-0.12} & \theta_{eq} = 60° \\ Oh^{-0.04} & \theta_{eq} = 150° \end{cases}. \tag{30}$$

From Eq. (30), the relationship between the asymmetry coefficient $L^*$ and $Oh$ can be expressed as:

$$L^* = \frac{L_{az,max}}{L_{ax,max}} \sim \frac{\omega_m^*}{L_m^*} \sim \begin{cases} Oh^{-0.04} & \theta_{eq} = 60° \\ Oh^{-0.02} & \theta_{eq} = 150° \end{cases}. \tag{31}$$

Figure 13(e) shows that a higher $Oh$ results in a modest decrease in the asymmetry of droplet spreading, and the effect is relatively small. Meanwhile, Figs. 13(c) and 13(d) demonstrate that the dimensionless times to reach the maximum axial spreading length $t_{ax,max}/t_\sigma$ and the maximum azimuthal spreading angle $t_{az,max}/t_\sigma$ are not affected by changes in $Oh$. However, a larger $Oh$ increases the droplet contact time, because during the retraction process, a higher $Oh$ will inhibit the retraction speed driven by surface tension, resulting in a longer contact time.

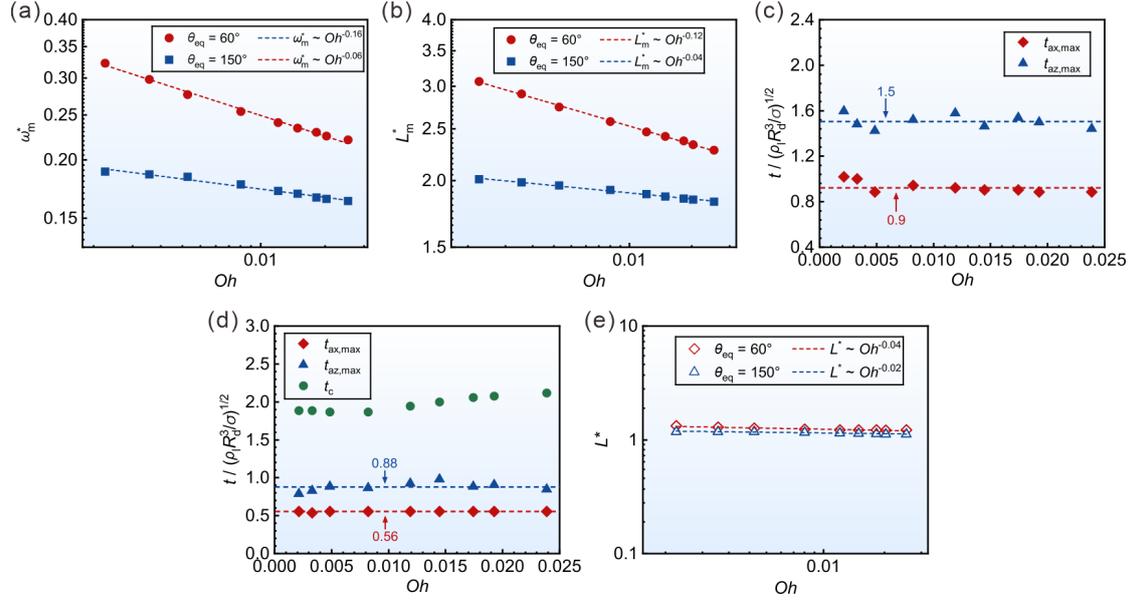

FIG. 13. Effect of $Oh$ on the azimuthal and axial spreading of droplets impacting a cylinder: (a) dependence of the dimensionless maximum azimuthal spreading angle $\omega_m^*$ on $Oh$; (b) dependence of the dimensionless maximum axial spreading length $L_m^*$ on $Oh$; (c) variation of the dimensionless characteristic times $t_{ax,max}/t_\sigma$ and $t_{az,max}/t_\sigma$ with $Oh$ in the deposition regime ($\theta_{eq} = 60°$); (d) variation of the dimensionless characteristic times $t_{ax,max}/t_\sigma$, $t_{az,max}/t_\sigma$ and $t_c/t_\sigma$ with $Oh$ in the rebound regime ($\theta_{eq} = 150°$); and (e) variation of the asymmetry coefficient $L^*$ with $Oh$. Here, $0.00213 \leq Oh \leq 0.0193$, $v_i = 1.0$ m/s, $We = 34.6$, and $D^* = 4.0$.

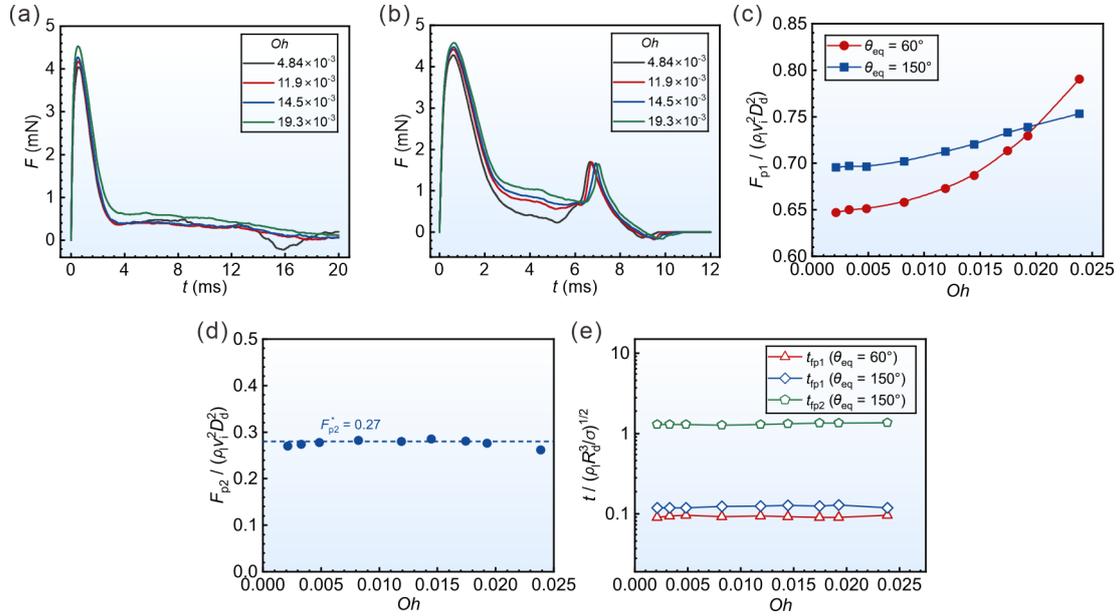

FIG. 14. Effect of $Oh$ on the impact force of droplets impinging a cylinder: (a) evolution of impact force under deposition regime ($\theta_{eq} = 60°$); (b) evolution of impact force under rebound regime ($\theta_{eq} = 150°$); (c) dependence of the first dimensionless peak impact force $F_{p1}^*$ on $Oh$; (d) dependence of the second dimensionless peak impact force

$F_{p2}^*$ on $Oh$; and (e) variation of the dimensionless characteristic times $t_{fp1}/t_\sigma$ and $t_{fp2}/t_\sigma$ with $Oh$. Here, $0.00213 \leq Oh \leq 0.0193$, $v_i = 1.0$ m/s, $We = 34.6$, and $D^* = 4.0$.

Finally, the influence of $Oh$ on the impact force was investigated. The evolution of the impact force with time at different $Oh$ is presented in Figs. 14(a) for $\theta_{eq} = 60°$ and 14(b) for $\theta_{eq} = 150°$. The first dimensionless peak impact force $F_{p1}^*$ increases with $Oh$, while the second dimensionless peak $F_{p2}^*$ remains almost unaffected, as depicted in Figs. 14(c) and 14(d). This is because a higher $Oh$ enhances viscous resistance, making droplet deformation more difficult during impact. Consequently, the droplet spreads more slowly in the early stage, and a larger part of its momentum is directly transferred to the wall within a short time, leading to an increase in the transient peak impact force. Within the investigated $Oh$ range in this study, the rebound regime remains unchanged, exhibiting a wing-like rebound. Since the rebound is primarily governed by inertial and capillary forces, the increased viscous dissipation associated with higher $Oh$ affects the retraction velocity and contact time but does not significantly alter the jump-off force. As presented in Fig. 14(e), the dimensionless time to reach the first peak $t_{fp1}/t_\sigma$ is insensitive to variations in $Oh$, whereas the dimensionless time to reach the second peak $t_{fp2}/t_\sigma$ slightly increases with $Oh$. This is because the first peak is caused by the sudden change in the initial momentum of the droplet, and its timescale is not affected by $Oh$. The increase in $Oh$ will slow down the speed of droplet retraction, resulting in a delay in the time of the second peak.

**E. Effect of diameter ratio**

The diameter ratio significantly affects the droplet impact process. To investigate this influence, numerical simulations are conducted using cylinders of various diameters. The value of $D^*$ is varied by changing the cylinder diameter, while the flat surface corresponds to an infinite diameter ratio $D^* \to \infty$. The findings reveal that the droplet achieves a larger maximum spreading area on a hydrophilic cylinder than on a flat wall; however, the spreading area is only slightly affected by the diameter ratio. For hydrophobic surfaces, although the diameter ratio has a minor effect on the spreading area, it significantly reduces the contact time during droplet impact, as shown in Figs. 15(a) and 15(b). A decrease in $D^*$ causes more mass and momentum of the droplet to transfer from the axial to the azimuthal direction during impact, resulting in an increase in the asymmetry coefficient $L^*$ and a reduction in the contact time on the hydrophobic wall, as illustrated in Fig. 15(e). The effects of $D^*$ on the dimensionless maximum spreading angle $\omega_m^*$ in the azimuthal direction and the dimensionless maximum spreading length $L_m^*$ in the axial direction are presented in Figs. 15(c) and 15(d), respectively. As the diameter ratio increases, $L_m^*$ increases and $\omega_m^*$ decreases significantly.

The variation of the impact force over time is presented for droplet impacts on hydrophilic and hydrophobic cylinders with different diameter ratios in Figs. 16(a) and 16(b). The variation of the dimensionless peak impact force with diameter ratio is illustrated in Fig. 16(c). Compared with the flat surface, the first dimensionless peak impact force decreases when the droplet impacts a cylinder and shows a further reduction as $D^*$ decreases. During the impact on the cylinder, the fluid flows around the circumference, retaining part of its momentum in the vertical direction. A larger curvature leads to greater retained momentum and consequently a smaller impact force. The second peak of the impact force is only slightly affected by $D^*$ on hydrophobic surfaces.

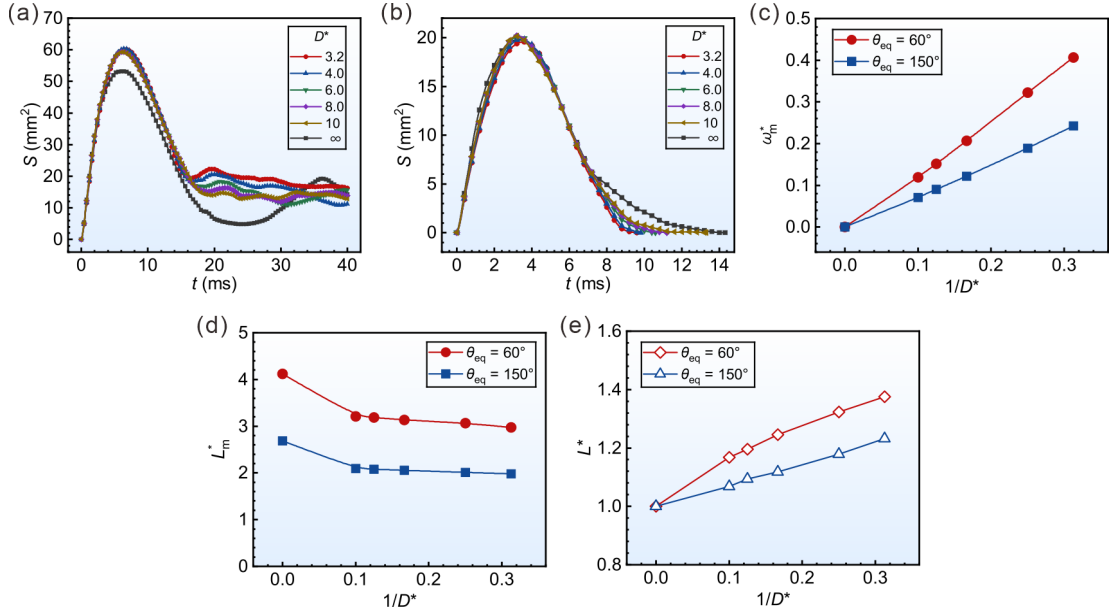

FIG. 15. Effect of $D^*$ on the kinematic parameters of droplet impact on a cylinder: (a) evolution of the spreading area for impact on a hydrophilic wall ($\theta_{eq} = 60°$); (b) evolution of the spreading area for impact on a hydrophobic wall ($\theta_{eq} = 150°$); (c) dependence of $D^*$ on the dimensionless maximum spreading angle $\omega_m^*$ in the azimuthal direction; (d) dependence of diameter ratio on the dimensionless maximum spreading length $L_m^*$ in the axial direction; and (e) dependence of diameter ratio on the asymmetry coefficient $L^*$. Here, $D_d = 2.5$ mm, $v_i = 1.0$ m/s, $\mu_l = 0.898$ mPa·s, $We = 34.6$, and $Oh = 0.00213$.

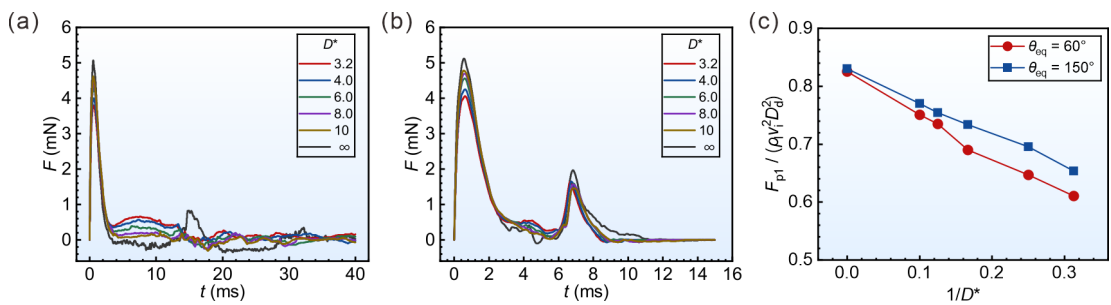

FIG. 16. Effect of $D^*$ on the impact force of droplet impact on a cylinder: (a) evolution of the impact force for impact on a hydrophilic wall ($\theta_{eq} = 60°$); (b) evolution of the impact force for impact on a hydrophobic wall ($\theta_{eq} = 150°$); (c) dependence of diameter ratio on $F_{p1}^*$. Here, $D_d = 2.5$ mm, $v_i = 1.0$ m/s, $\mu_l = 0.898$ mPa·s, $We = 34.6$, $Oh = 0.00213$.

## IV. Conclusions

In the present study, the kinematic and dynamic characteristics of droplet impact on a cylinder are studied by numerical simulation. The effects of surface wettability, Weber number, Ohnesorge number, and diameter ratio on the characteristic parameters are analyzed. The results show that there are two typical impact regimes, including deposition and rebound, and each of them consists of three different stages. In the deposition regime, the impact force curve exhibits a single peak, whereas in the rebound regime, two distinct peaks appear. As the contact angle increases, the dimensionless maximum spreading area, maximum spreading angle, and maximum spreading length all decrease. Meanwhile, the asymmetry coefficient gradually decreases and eventually stabilizes at 1.17. Wettability has little influence on the impact force during the initial spreading stage but significantly affects the second peak in the rebound stage. The dimensionless time corresponding to the second peak and the contact time both exhibit a linear relationship with $(1-\cos\theta_{eq})^{-1/2}$. With increasing *We*, the dimensionless maximum spreading area, dimensionless time to reach maximum spreading, dimensionless maximum spreading angle, dimensionless maximum spreading length, and asymmetry coefficient all increase and follow power-law correlations. The Weber number does not significantly affect the overall trend of the impact force but strongly influences the magnitude and timing of the peak forces. As *We* increases, the first dimensionless peak impact force tends to stabilize at 0.63, while the second dimensionless peak impact force tends to be stable when *We* is large, but fluctuates greatly when *We* is small. In addition, the dimensionless maximum spreading area, dimensionless spreading angle, dimensionless spreading length, and asymmetry coefficient decrease with increasing Ohnesorge number, also following power-law relationships. The first dimensionless peak impact force increases with *Oh*, whereas the second peak remains nearly unaffected. Increasing the diameter ratio leads to a higher first dimensionless peak impact force. Overall, the results of this study provide deeper insight into the kinematic and dynamic characteristics of droplet impact on cylinders and offer valuable guidance for optimizing industrial processes, designing functional materials, and advancing research in multiphase flow dynamics.

## Acknowledgements

This work was supported by the National Natural Science Foundation of China (Grant No. U25A20191).